\documentclass[aps,pra,twocolumn,groupedaddress,showpacs]{revtex4-1} %add ',preprint' after 'pra' to make double spaced

\usepackage{graphicx} %this is needed to include graphics (png and pdf)
\usepackage{hyperref}
\usepackage{textcomp}
\begin{document}

%\preprint{}

\title{All-Optical Broadband Excitation of the Motional State of Trapped Ions}


\author{Kevin Sheridan, Nicolas Seymour-Smith, Amy Gardner and Matthias Keller}
%\author{}
%\homepage[]{http://www.itcq.org.uk/}
%\thanks{}
\affiliation{Molecular Physics Laboratory, Department of Physics and Astronomy, University of Sussex, Falmer, BN1 9QH, England.}
\email[]{K.T.Sheridan@sussex.ac.uk} 

\date{\today}

\begin{abstract}
We have developed a novel all-optical broadband scheme for exciting, amplifying and measuring the secular motion of ions in a radio frequency trap. Oscillation induced by optical excitation has been coherently amplified to precisely control and measure the ion's secular motion. Requiring only laser line-of-sight, we have shown that the ion's oscillation amplitude can be precisely controlled. Our excitation scheme can generate coherent motion which is robust against variations in the secular frequency. Therefore, our scheme is ideal to excite the desired level of oscillatory motion under conditions where the secular frequency is evolving in time. Measuring the oscillation amplitude through Doppler velocimetry, we have characterized the experimental parameters and compared them with a molecular dynamics simulation which provides a complete description of the system.
\end{abstract}

\pacs{32.80.-t}
%\keywords{}

\maketitle

%%INTRODUCTION%%
%%%%%%%%%%%%%%%%
\section{\label{sec:Intro}Introduction}
The optical excitation of the secular motion of trapped ions is an invaluable tool in modern quantum information processing \cite{Ekert,Kimble,Leibfried2}, high resolution spectroscopy \cite{Koelemeij} and quantum optics \cite{Bloch}. Precisely tailored laser pulses are employed to transfer the state of one ion to another \cite{Turchette}, entangle several ions \cite{Haffner,Blatt}, investigate the behavior of coupled systems \cite{Riebe} and generate coherent phonon fields \cite{Vahala}. Employing laser pulses on the motional sidebands, several quantum gates have been demonstrated \cite{Cirac} and ions have been entangled though their joint motion in the trapping potential \cite{Kim}. Laser induced dipole forces have been utilized to excite the coherent motion of ions which in turn has been exploited to demonstrate quantum phase gates and entangle multiple ions \cite{Leibfried3,Biercuk2}. Recently, the laser induced coupling between the internal and external degrees of freedom of trapped ions has been applied to high resolution spectroscopy of ions which are not suitable for electron shelving \cite{Schmidt}. 

In all these applications, precise control and measurement of the ion's secular motion is required. Several schemes based on the electric excitation of the ion's motion have been developed to precisely measure the secular frequency \cite{Dholakia,Welling,Schlemmer,Sheridan}. In addition, the ion's secular motion can be excited and measured using narrowband excitation through the modulation of the interaction laser intensity \cite{Drewsen}. In this article we report a novel scheme to optically excite and control the amplitude of the ion's secular motion through broadband excitation. Employing a fast change in radiation pressure, an oscillation of the trapped ion around the new, shifted equilibrium position is induced which in turn is amplified by blue detuning the interaction laser. Subsequently, the ion's oscillation is measured through Doppler velocimetry \cite{Blumel,Berkeland,Biercuk}. The scheme is fast and doesn't require stringent control of the laser parameters. Furthermore, since a coherent excitation of the ion's motion is not required, the method is well suited for situations in which the secular frequency is unknown or rapidly changing. 
%\cite{Hume}

In the first section of this article, the principle of our measurement scheme is discussed. This is followed by a description of the experimental methodology and set-up. Measurements, which have been performed to demonstrate and characterize the optical excitation and amplification of the ion using intensity switching excitation, are then presented. A molecular dynamics simulation, which takes into account all of the relevant parameters of the system, is used to generate fitting curves for the experimental data. Finally, the application of frequency switching excitation to precision measurement of the ion's time varying secular frequency is discussed.

%%%%%%%%%%%%%%%%%%%%%%%%%%%%%%%%%%%%%%%%%%%%%%%%%%%%%%%%%%%%%%%%%%%%%%%%%%%%%%%%%%%%%%%%%%%%%%%%%%%%%%%%%%%%%%%%%%%%%%%%%%%%%%%%%%%%%%%%%%
%%MEASUREMENT PRINCIPLE%%
%%%%%%%%%%%%%%%%
\section{\label{sec:measurement_principle}Measurement principle}
Our technique is based on a rapid change in the magnitude of the laser induced radiation pressure. The ion is trapped in a linear radio frequency (rf) ion-trap and laser cooled with a laser beam aligned along the trap axis. Starting with the laser cooled ion in equilibrium, a fast change in the laser intensity, or detuning, associated with a change in the ion's fluorescence rate $\Delta F$ results in an axial displacement of the ion in the trapping potential due to the altered radiation pressure. Consequently, the ion oscillates with the secular frequency $\omega_{0}$ around the new equilibrium position. To detect the ion's oscillation, Doppler velocimetry is utilized and the ion's oscillatory motion is mapped onto its fluorescence. The ion's oscillation amplitude can be expressed in terms of the maximal Doppler shift $\Delta$ with
\begin{equation}
\Delta =\frac{2\pi h \Delta F}{m\omega_{0}\lambda^{2}},
\label{equ:osc_amp}
\end{equation}
where $m$ is the mass of the ion, $\lambda$ is the interaction laser wavelength and $h$ is Planck's constant. Thus, the ion's oscillation amplitude is proportional to the magnitude of the change in fluorescence rate.

The ion's oscillation amplitude is amplified by blue detuning the cooling laser \cite{Hume,Vahala}. For a blue detuned laser, the radiation pressure for co-propagating ions exceeds that of the counter-propagating case. This results in an enhancement of the ion's oscillation velocity and thus its amplitude. The increase in modulation amplitude can, for small amplitudes, be described by an exponential increase with amplification time $\tau$ by
\begin{equation}
G=\mathrm{exp}\left[\frac{\pi h}{m\lambda^{2}}\left.\frac{dF}{d\delta}\right|_{\delta_{0}} \tau \right],
\label{equ:amp_gain}
\end{equation}
where $F(\delta)$ is the ion's fluorescence rate and $\delta_{0}$ is the detuning of the interaction laser. Even though this describes the behavior of the coherent motion well, the blue detuning of the laser also induces heating which in turn results in a motional broadening of the fluorescence spectrum. This broadening causes a significant decrease in the gain of the coherent motion. In order to include the laser heating in the measurement analysis, we employ a molecular dynamics simulation (discussed in section \ref{sec:numerical_simulation}), which takes into account the recoil of photon scattering events.

The oscillation of the ion's velocity results in a periodic change in the laser detuning, in the reference frame of the ion, due to the linear Doppler effect. The resulting fluorescence modulation $\Delta F_{m}$ can be expressed for small velocities as
\begin{equation}
\Delta F_{m} = \frac{2\pi h \Delta F}{m\omega_{0}\lambda^{2}}\left.\frac{dF}{d\delta}\right|_{\delta_{0}} G.
\label{equ:F_mod}
\end{equation}
Thus, the transfer of the Doppler induced detuning modulation depends on the gradient of the fluorescence spectrum. The detection efficiency can therefore be optimized by tuning the detection laser to the point of steepest slope on the spectral lineshape.

The interaction with a red detuned laser results in a damping of the coherent oscillation accompanied by a cooling of the ion's thermal motion. Consequently, the coherent motion ceases and the ion's kinetic energy approaches the Doppler cooling limit. 

In order to measure the ion's time dependant fluorescence rate (fluorescence profile), the ion fluorescence detection times are continuously recorded and correlated with the excitation pulse. To analyze the measurements, we fit the results of our simulation to the experimentally obtained fluorescence profile. From the fit we can determine the oscillation amplitude of the ion and the induced heating.

%%%%%%%%%%%%%%%%%%%%%%%%%%%%%%%%%%%%%%%%%%%%%%%%%%%%%%%%%%%%%%%%%%%%%%%%%%%%%%%%%%%%%%%%%%%%%%%%%%%%%%%%%%%%%%%%%%%%%%%%%%%%%%%%%%%%%%%%%%
%%EXPERIMENTAL SETUP%%
%%%%%%%%%%%%%%%%
\section{\label{sec:exp_set_up}Experimental setup}
\begin{figure}[h]
\begin{center}
\resizebox{0.45\textwidth}{!}{\includegraphics{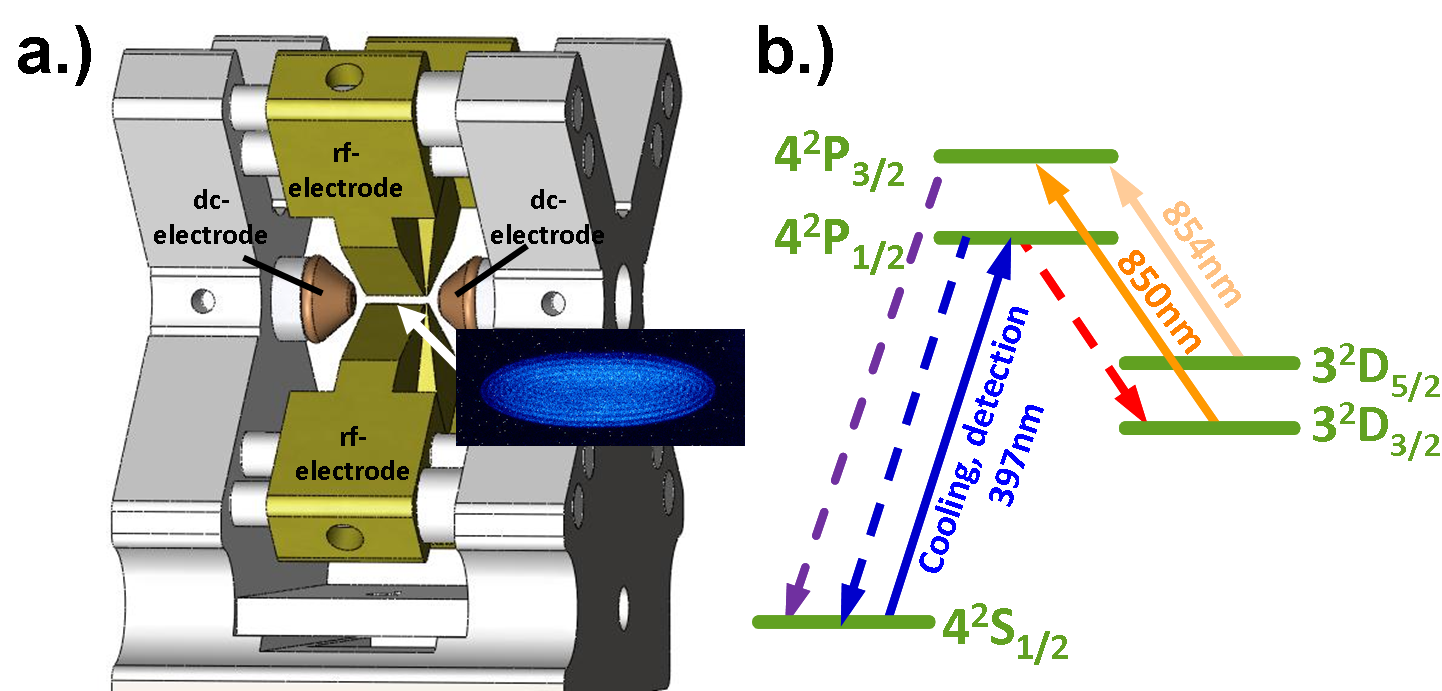}}
\caption[]{(a) Schematic of the ion trap. (b) Level scheme of $^{40}$Ca-ion including all lasers required to cool and re-pump the ion (solid arrows). Dashed lines indicate spontaneous emission.}
\label{fig:iontrap}
\end{center}
\end{figure}

The trap used in this experiment is a linear rf Paul trap. A schematic of the ion trap is shown in figure \ref{fig:iontrap}(a). It consists of four blade shaped rf-electrodes, which provide the radial confinement of the ions. The ion-electrode separation is 465~$\mu$m and the rf-electrode length is 4~mm. Positive static potentials applied to two dc-electrodes are used for the axial ion confinement. Each of the dc-electrodes has an opening in order to provide laser access to the trapped ions exactly along the trap axis.
The trap is operated at a frequency of 26~MHz and a typical rf-power of 300~mW to obtain a radial secular frequency of 1.5~MHz. The typical axial confinement is 250~kHz at a dc-voltage of 200~V.

The $^{40}$Ca-ion is Doppler-cooled on the 4S$_{1/2}\!\!\rightarrow $4P$_{1/2}$ transition at 397~nm (see figure \ref{fig:iontrap}(b)) by two laser beams, one aligned radially  and one aligned along the trap axis. Consequently, all directions of the ion's motion are cooled. In order to avoid populating the meta-stable D$_{3/2}$-state through the decay of the P$_{1/2}$-level, a re-pump laser on the D$_{3/2}\!\!\rightarrow $P$_{3/2}$ transition at 850~nm is applied. From here, the ion is returned to the ground state through the spontaneous decay of the P$_{3/2}$-state. However, another re-pump laser at 854~nm must be applied in order to avoid optical pumping into the meta-stable D$_{5/2}$-state. This re-pump scheme has the advantage over re-pumping via the P$_{1/2}$-level in that it creates an effective two-level cooling system without coherence effects from coupling of the cooling and re-pump lasers.

The trap is loaded through the photoionization of neutral calcium atoms effusing from a resistively heated oven located below the trap and aligned perpendicular to the trap axis. Photoionization is a two-photon process. The first photon is resonant with the S$_{0}\!\!\rightarrow$P$_{1}$ transition of neutral calcium at 423~nm and the second photon at 375~nm provides the energy necessary to ionize the excited state calcium atoms \cite{Lucas}. For appropriate trapping parameters, single ions, linear ion strings and large three-dimensional (3D) Coulomb crystals with several hundred ions can be quickly and repeatably loaded into the trap.

The cooling and re-pump laser beams are individually controlled by AOMs in double pass configuration and delivered to the experiment through polarization maintaining fibres. This set-up allows for precise experimental control of the laser frequency as well as the laser intensity. The AOMs are controlled by either of two computer controlled voltage-controlled oscillators (VCOs) which are connected to the AOM through rf-switches. Additionally, a computer controlled pulse generator determines the timing sequences for switching the laser intensity and for switching between the two VCOs via the rf-switch. In this way, the frequency and intensity of the 397~nm laser may be instantly shifted to pre-selected values whenever required.

The ion's fluorescence at 397~nm is collected by a microscope lens with a numerical aperture of 0.22. During a measurement, the photon detection times are recorded with a time to digital converter (FastComtec P7888) and correlated with a trigger pulse occurring at the beginning of each measurement sequence. After the chosen measurement duration, the photon counts are binned according to arrival time and the counts for all measurement sequences are summed. The motional spectrum of the ion is obtained by taking the fast Fourier transform (FFT) of the fluorescence profile allowing the center-of-mass (COM) mode frequency to be measured with high precision. A fitting curve is generated from the simulation and the ion's peak oscillation amplitude and measurement induced temperature increase are determined. 

%%%%%%%%%%%%%%%%%%%%%%%%%%%%%%%%%%%%%%%%%%%%%%%%%%%%%%%%%%%%%%%%%%%%%%%%%%%%%%%%%%%%%%%%%%%%%%%%%%%%%%%%%%%%%%%%%%%%%%%%%%%%%%%%%%%%%%%%%%
%%LASER INTENSITY SWITCHING%%
%%%%%%%%%%%%%%%%
\section{\label{sec:exp_procedure}Laser intensity switching}
The experimental sequence we employ begins with red detuned lasers for Doppler cooling the ion. This prepares the ion in a motional state close to the Doppler cooling limit. After this initial cooling period, the ion's equilibrium position in the trapping potential is shifted by a change in the laser induced radiation pressure. When the laser is switched off to impart the motional excitation, the ion undergoes undamped oscillations at the axial secular frequency. Subsequently applying a blue detuned laser, the ion's motion is coherently amplified. Then, switching the laser back to red detuning damps the ion's motion. This sequence is repeated at a rate of 1~kHz. To obtain a fluorescence profile during the measurement, a histogram of the photon arrival times with respect to each trigger pulse is generated. The photon arrival times for each sequence are summed to form a total fluorescence profile.

\begin{figure}[h]
\begin{center}
\resizebox{0.475\textwidth}{!}{\includegraphics{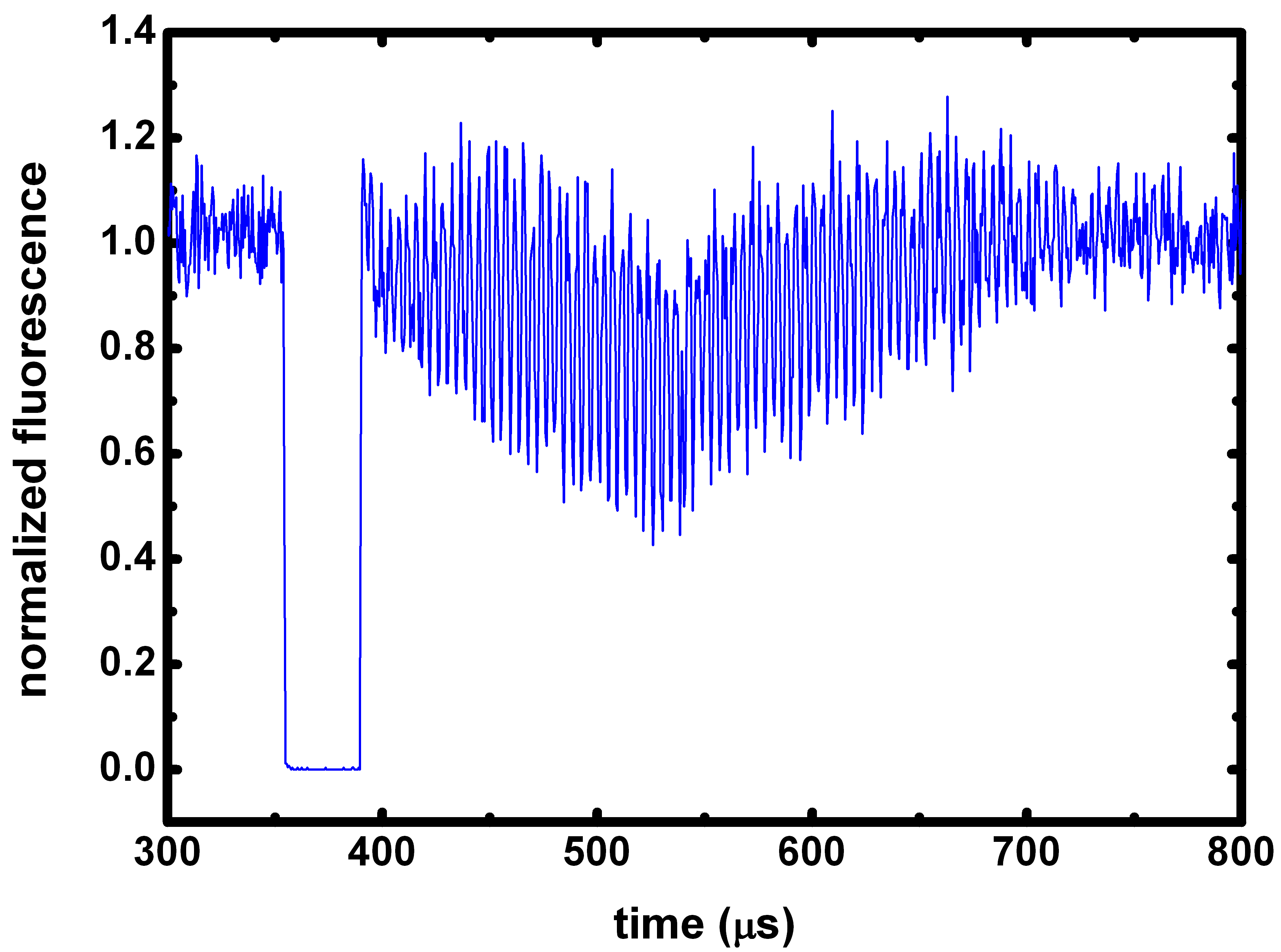}}
\caption{Fluorescence as a function of time elapsed since the experimental trigger produced using the laser intensity switching technique. The laser detuning is +10~MHz during amplification and -10~MHz during re-cooling. The laser intensity is 2$I_{sat}$.}
\label{fig:exp_kick_demo}
\end{center}
\end{figure}

A typical fluorescence profile for a measurement time of ten seconds is shown in figure \ref{fig:exp_kick_demo}. In the initial 355~$\mu$s, the laser is red detuned to cool the ion's motion. This is followed by a 35~$\mu$s interruption in the laser intensity giving a drop in the ion's fluorescence signal and a corresponding change in the laser induced radiation pressure. The ion's oscillation, resulting from the sudden shift in equilibrium position, is then amplified by switching the laser on with blue detuning. The increase in the amplitude of coherent oscillatory motion due to the interaction with the blue detuned laser is clearly visible. In addition, a decrease in the mean signal level is observed. The change in the mean signal level is a measure of the increased linewidth of the heated ion's fluorescence spectrum and therefore an indicator of the measurement induced temperature increase (see section \ref{sec:blue-tuning}). After an amplification time of 150~$\mu$s, the laser is red detuned in order to re-cool the ion. This results in a damping of the coherent motion which ceases after a further 150~$\mu$s. In addition, the mean signal level and the temperature of the ion return to their equilibrium values during the re-cooling period.

%%%%%%%%%%%%%%%%%%%%%%%%%%%%%%%%%%%%%%%%%%%%%%%%%%%%%%%%%%%%%%%%%%%%%%%%%%%%%%%%%%%%%%%%%%%%%%%%%%%%%%%%%%%%%%%%%%%%%%%%%%%%%%%%%%%%%%%%%%
%%MOLECULAR DYNAMICS SIMULATION%%
%%%%%%%%%%%%%%%%
\subsection{\label{sec:numerical_simulation}Molecular dynamics simulation}
\begin{figure}[h]
\begin{center}
\resizebox{0.475\textwidth}{!}{\includegraphics{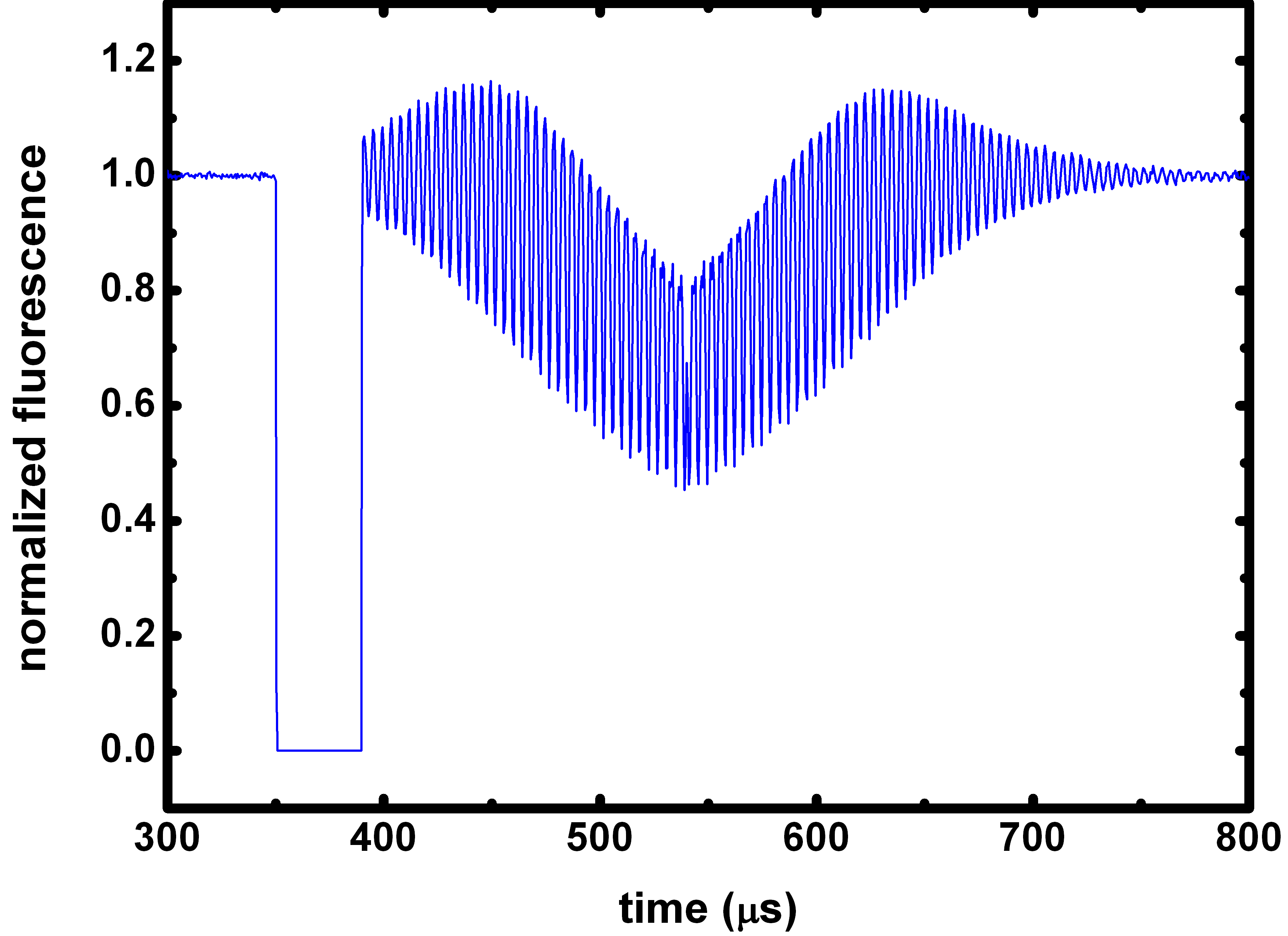}}
\caption{Numerical simulation of the experimentally measured fluorescence profile shown in figure \ref{fig:exp_kick_demo} using the same parameters as those used in the experiment.}
\label{fig:sim_kick_demo}
\end{center}
\end{figure}

We can reproduce the behavior of this system well with a molecular dynamics simulation which we have developed. The simulation takes into account all of the relevant parameters of the system. For each step in the simulation, the ion's interaction with the laser is used to determine the population of the excited state in the two-level approximation. The probability of photon absorption and subsequent spontaneous emission is calculated from this, and used to determine the time between spontaneous emission events. The momentum imparted to the ion by absorption and spontaneous emission is used to calculate and update its position and velocity. Finally, photons emitted by the ion are counted to form a fluorescence profile. The laser pulse sequence is chosen to match the experiment with corresponding excitation, amplification and re-cooling timing. Figure \ref{fig:sim_kick_demo} shows the fluorescence profile produced by the simulation for the same parameters used to produce the experimental data in figure \ref{fig:exp_kick_demo}. The agreement between the fluorescence profile created by the simulation and the experimental result is very good.

%%Amplification time%%
\subsection{\label{sec:blue-tuning}Amplification time}
\begin{figure}[h]
\begin{center}
\resizebox{0.475\textwidth}{!}{\includegraphics{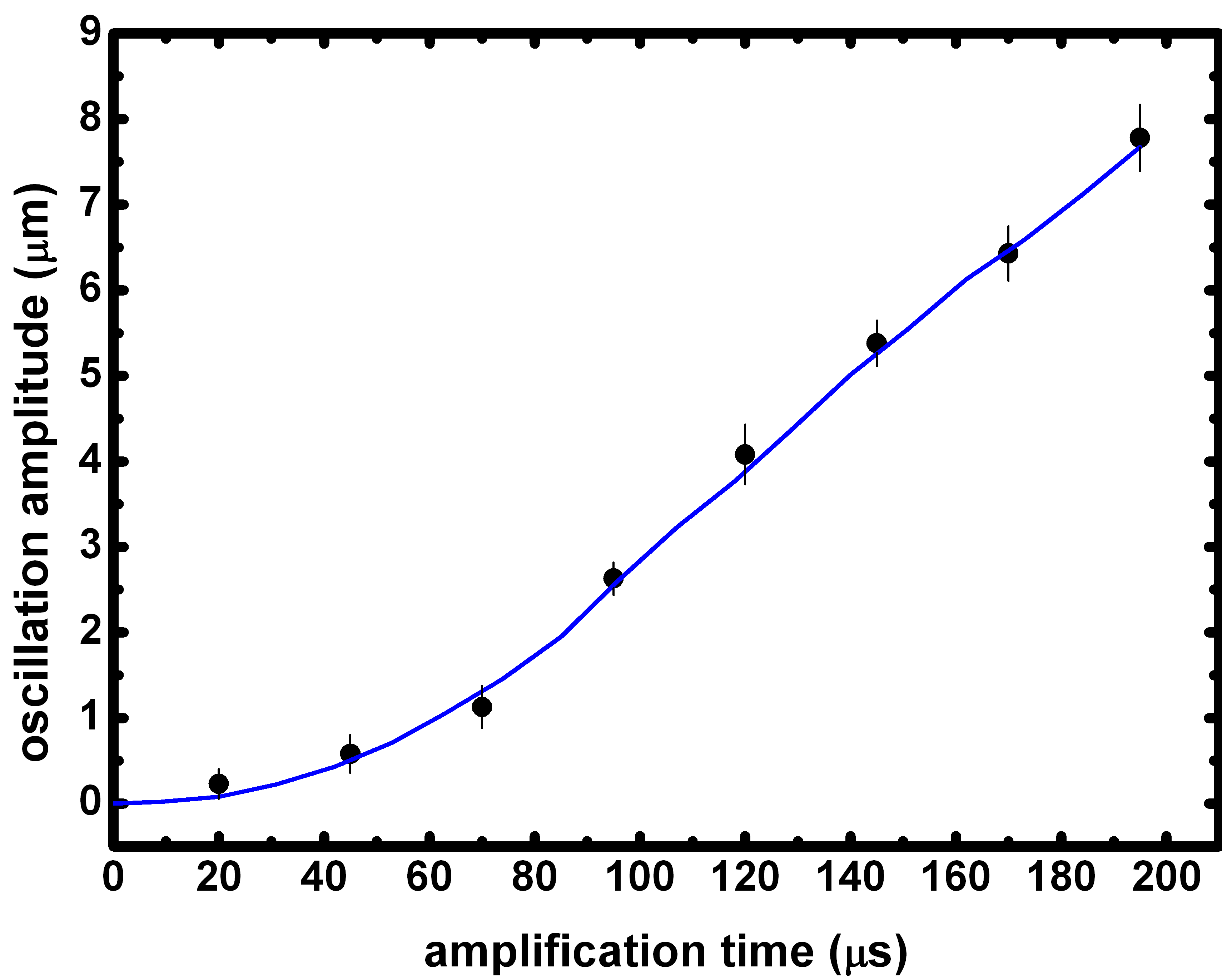}}
\caption{The ion's oscillation amplitude as a function of the amplification time. For each measurement The laser detuning is +10~MHz during amplification and -10~MHz during re-cooling. The laser intensity is 2$I_{sat}$. A molecular dynamics simulation has been fit to the data (solid line). The relationship is initially exponential but becomes approximately linear for larger amplification times.}
\label{fig:exp_vary_blue_time}
\end{center}
\end{figure}
Figure \ref{fig:exp_vary_blue_time} shows the ion oscillation amplitude as a function of the duration of laser amplification. For each amplification time, a simulated fluorescence profile is fit to the experimental data. The oscillation amplitude is then determined from the simulation using the best fit parameters. For small amplification times, the relationship between the ion oscillation amplitude and the amplification time is approximately exponential, as described by equation \ref{equ:amp_gain}. However, for longer amplification times the relationship becomes nearly linear. This is due to the heating of the ion and associated broadening of the ion's spectral line. The increased linewidth of the heated ion decreases the slope of the spectral lineshape and in turn decreases the amplification efficiency.

From the measured fluorescence profile it is possible to unambiguously distinguish between the ion's thermal motion and its coherent oscillatory motion. During amplification of the ion's motion, Doppler broadening leads to an increase in the inhomogeneous component of the ion's spectral line. For an ion with a temperature of $T$, the inhomogeneous linewidth contribution is given by
\begin{equation}
\sigma = \frac{2\pi}{\lambda}\sqrt{\frac{k_{B}T}{m}}.
\label{equ:inhomo_linewidth}
\end{equation} 
The heated ion's fluorescence spectrum can be described by a Voigt profile. For a fixed laser detuning and small amplitude oscillation, the discrepancy between the mean fluorescence level at the start of amplification and at the end is caused by the increased inhomogeneous contribution to the ion's linewidth. For small blue detuning, the increasing linewidth results in a decreasing mean fluorescence level, as is demonstrated in figure \ref{fig:exp_kick_demo}. However, for large blue detuning, the increasing linewidth will result in an increase in the fluorescence level (see, for example, the slight increase in mean signal level during the amplification in figure \ref{fig:exp_detuning_shift_kick}).  
\begin{figure}[h]
\begin{center}
\resizebox{0.475\textwidth}{!}{\includegraphics{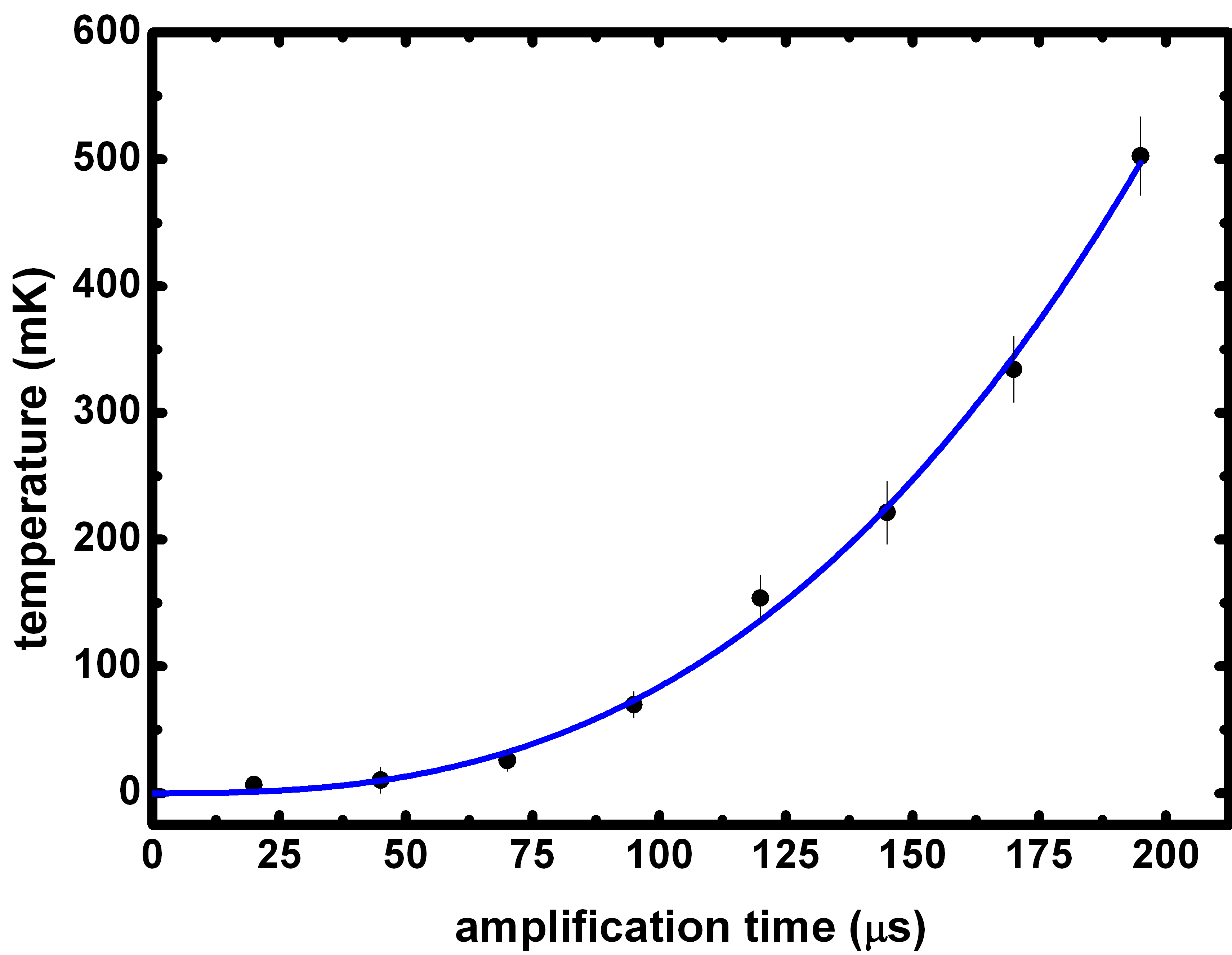}}
\caption{The measurement induced temperature increase of the ion as a function of the amplification time. Each data point corresponds to a point in figure \ref{fig:exp_vary_blue_time}. The expected quadratic function has been fit to the data.}
\label{fig:exp_ion_temp}
\end{center}
\end{figure}

We measure the fluorescence spectrum of the Doppler cooled ion prior to each measurement in order to determine the homogeneous linewidth. The fluorescence spectrum of the cooled ion is a Lorentzian function with an estimated temperature on the order of the Doppler cooling limit. From the change in mean signal level following amplification we can then determine the measurement induced inhomogeneous contribution to the total linewidth. The increase in the ion temperature relative to its initial state is then calculated from the inverted equation \ref{equ:inhomo_linewidth} \cite{Brama,Wesenberg}.

The ion temperature at the end of amplification as a function of the amplification time is shown in figure \ref{fig:exp_ion_temp}. The relationship is approximately quadratic which reflects the approximately linear change in linewidth for increasing amplification time and the quadratic relationship between linewidth and temperature given by equation \ref{equ:inhomo_linewidth}. For amplification times beyond around 100~$\mu$s the measurement induced temperature increase exceeds 100~mK, which is on the order of the temperature required to form ion Coulomb crystals \cite{Hornekaer}. For amplification times below 50~$\mu$s the measurement induced temperature increase remains below 10~mK which is low enough to allow for the stable trapping of large 3D Coulomb crystals.

%%%%%%%%%%%%%%%%%%%%%%%%%%%%%%%%%%%%%%%%%%%%%%%%%%
%%Laser intensity%%
\subsection{\label{sec:laser_power}Laser intensity}
\begin{figure}[h]
\begin{center}
\resizebox{0.475\textwidth}{!}{\includegraphics{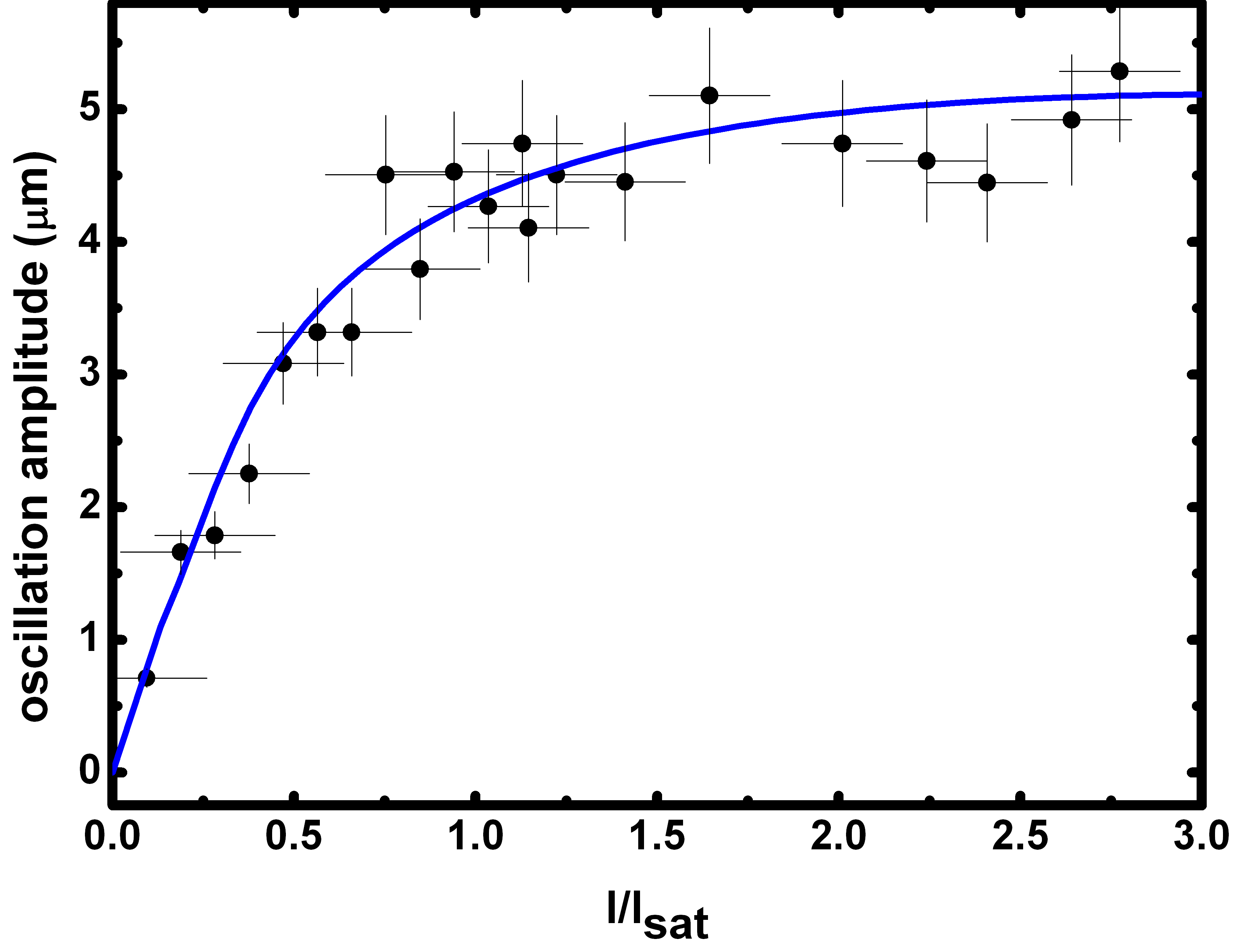}}
\caption{The ion's oscillation amplitude as a function of the laser intensity in units of the saturation intensity. For each measurement the laser detuning is +10~MHz during amplification and -10~MHz during re-cooling. The amplification time is 150~$\mu$s. A molecular dynamics simulation has been fit to the data (solid line).}
\label{fig:exp_vary_power}
\end{center}
\end{figure}
The laser intensity determines the excited state population of the ion, the width of the spectral line and the magnitude of the optical excitation. At low laser intensities, the ion's oscillation amplitude increases linearly with the magnitude of the change in the fluorescence rate $\Delta F$ resulting from excitation (see equation \ref{equ:F_mod}). As the intensity is increased further, saturation occurs resulting in a saturation of the fluorescence modulation signal and thus a decrease of the amplification gain. 

We have measured the dependence of the ion's oscillation amplitude on the laser intensity while keeping all other parameters constant. In order to determine the amplitude of the coherent motion, the ion's motional spectrum has been employed. The motional spectrum amplitude is measured by taking the FFT of the fluorescence profile. The oscillation amplitude is determined by fitting a simulated fluorescence profile to one experimental data point. A calibration coefficient relating the amplitude of the motional spectrum to the oscillation amplitude is then determined. A fitting curve for the experimentally measured motional spectrum amplitude as a function of the laser intensity is generated by the molecular dynamics simulation and the result is converted into units of oscillation amplitude using the calibration coefficient. The results are shown in figure \ref{fig:exp_vary_power}. 

%%Laser detuning%%
\subsection{\label{sec:laser_detuning}Laser detuning}
\begin{figure}[h]
\begin{center}
\resizebox{0.475\textwidth}{!}{\includegraphics{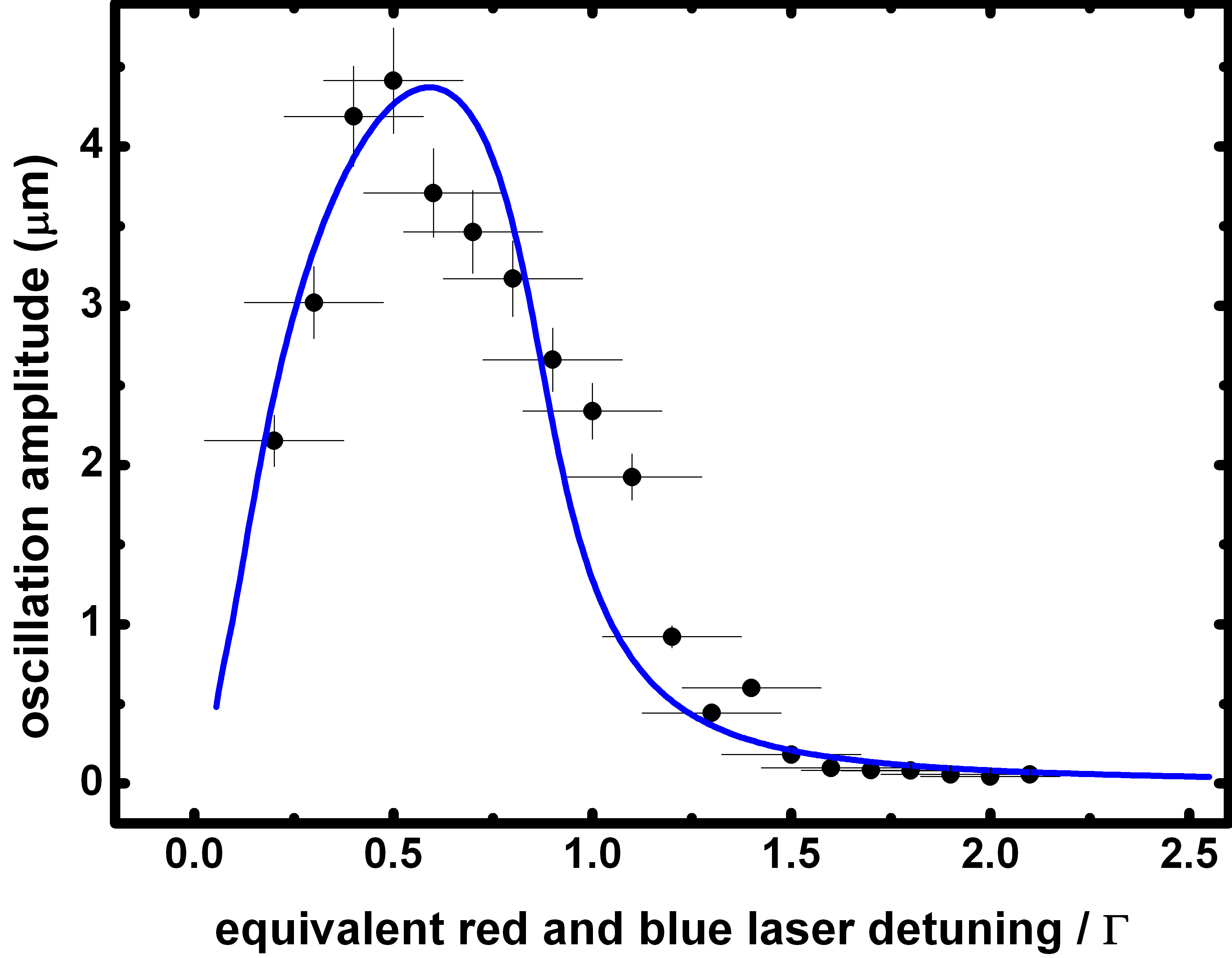}}
\caption{The ion's oscillation amplitude as a function of the laser detuning in units of the natural linewidth $\Gamma$. The x-axis is the equivalent blue and red laser detuning used during amplification and re-cooling respectively. For each measurement the laser intensity is 2$I_{sat}$ and the amplification time is 150~$\mu$s. A molecular dynamics simulation has been fit to the data (solid line).}
\label{fig:exp_vary_detuning}
\end{center}
\end{figure}
We have measured the ion's oscillation amplitude as a function of the laser detuning. The laser blue and red detuning are equal with respect to the center of the spectral line for each of the measurements. This ensures that a difference in laser detuning does not contribute to the optical excitation amplitude. The results are shown in figure \ref{fig:exp_vary_detuning}. A curve generated by the molecular dynamics simulation has been fit to the data. 

The effect of the laser detuning on the fluorescence modulation signal is partly determined by the slope of the ion's spectral line. The most efficient amplification of the ion's oscillatory motion is achieved for the laser tuning corresponding to the point of steepest slope. For detunings closer to the spectral line center the amplification decreases due to the decrease in the gradient. However, at the same time, the increase in total fluorescence, and corresponding increase in radiation pressure, for smaller laser detuning results in a shift of the optimum detuning towards the line center. 

%%Laser off-time%%
\subsection{\label{sec:laser_off_time}Laser off-time}
\begin{figure}[h]
\begin{center}
\resizebox{0.475\textwidth}{!}{\includegraphics{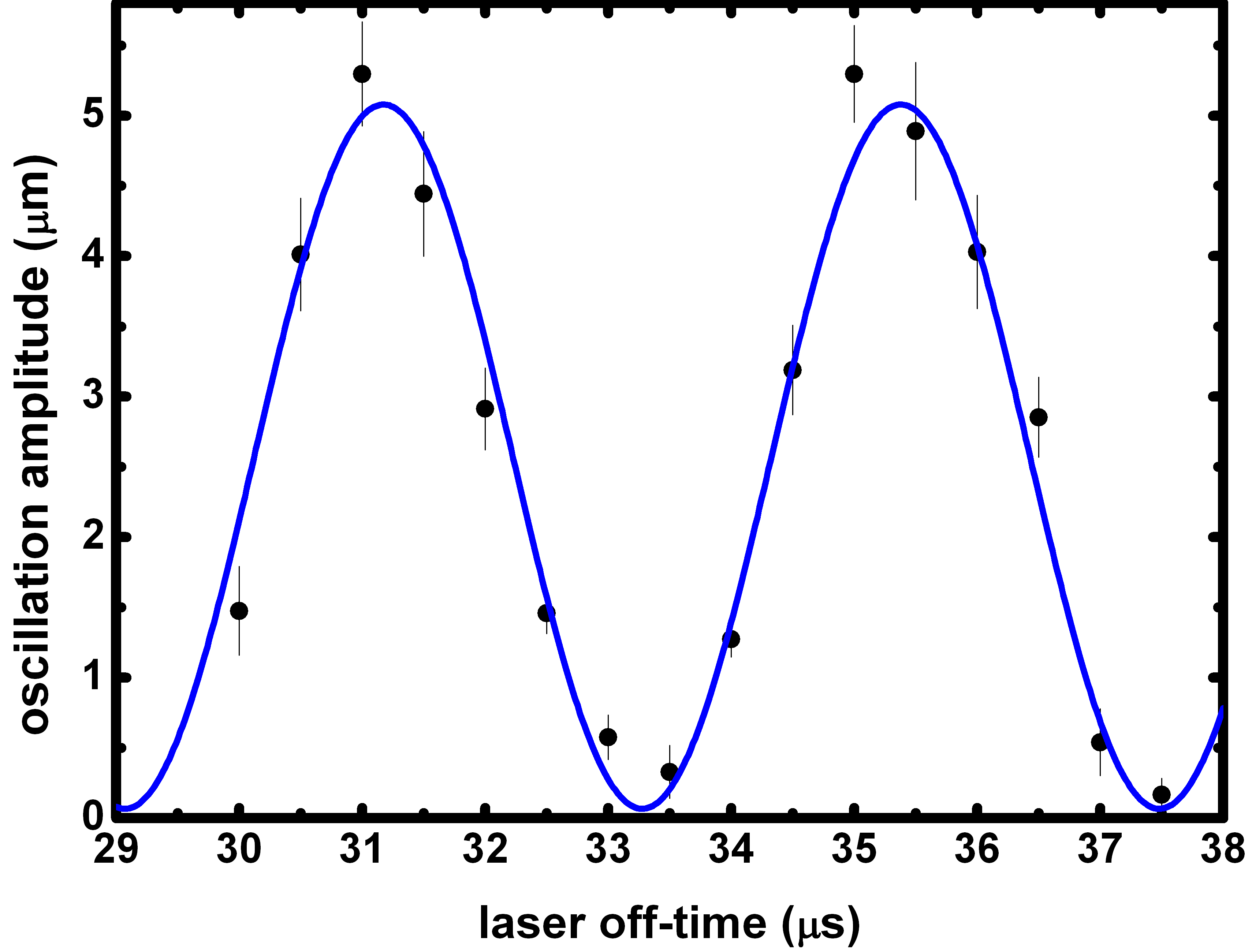}}
\caption{The ion's oscillation amplitude as a function of the laser off-time. For each measurement the laser detuning is +10~MHz during amplification and -10~MHz during re-cooling. The laser intensity is 2$I_{sat}$ and the amplification time is 150~$\mu$s. A sine-squared function has been fit to the data.}
\label{fig:exp_vary_off_time}
\end{center}
\end{figure}
When the interaction laser is switched off, the ion behaves as an undamped harmonic oscillator with oscillation amplitude determined by its displacement from the equilibrium position in the trap potential. For laser off-times equal to $n\!\cdot\!T$, where $T$ is the period of the ion's oscillation and $n$ is an integer, the ion's motion is exactly out-of-phase with the force from the laser switching back on. In this case, when the laser is reintroduced the ion will be located in the equilibrium position of the combined ion-trap and light pressure potential, therefore the ion's oscillatory motion will cease. If, however, the laser off-time is equal to (n+$\,$\textonehalf$\,) \! \cdot \! T$, the ion's motion will be in-phase with the laser and the momentum imparted through the optical excitation will enhance the coherent motion of the ion. We have measured this effect by varying the laser off-time while keeping all other parameters fixed. The result is shown in figure \ref{fig:exp_vary_off_time} where the ion's oscillation amplitude is plotted as a function of the laser off-time. The expected sine-squared function has been fit to the data. The dependance of the excitation amplitude on the switching time can be avoided through excitation by laser frequency switching while keeping the laser intensity constant.

%%%%%%%%%%%%%%%%%%%%%%%%%%%%%%%%%%%%%%%%%%%%%%%%%%%%%%%%%%%%%%%%%%%%%%%%%%%%%%%%%%%%%%%%%%%%%%%%%%%%%%%%%%%%%%%%%%%%%%%%%%%%%%%%%%%%%%%%%%
%%LASER FREQUENCY SWITCHING%%
%%%%%%%%%%%%%%%%
\section{\label{sec:detuning_kicking}Laser frequency switching}
\begin{figure}[h]
\begin{center}
\resizebox{0.475\textwidth}{!}{\includegraphics{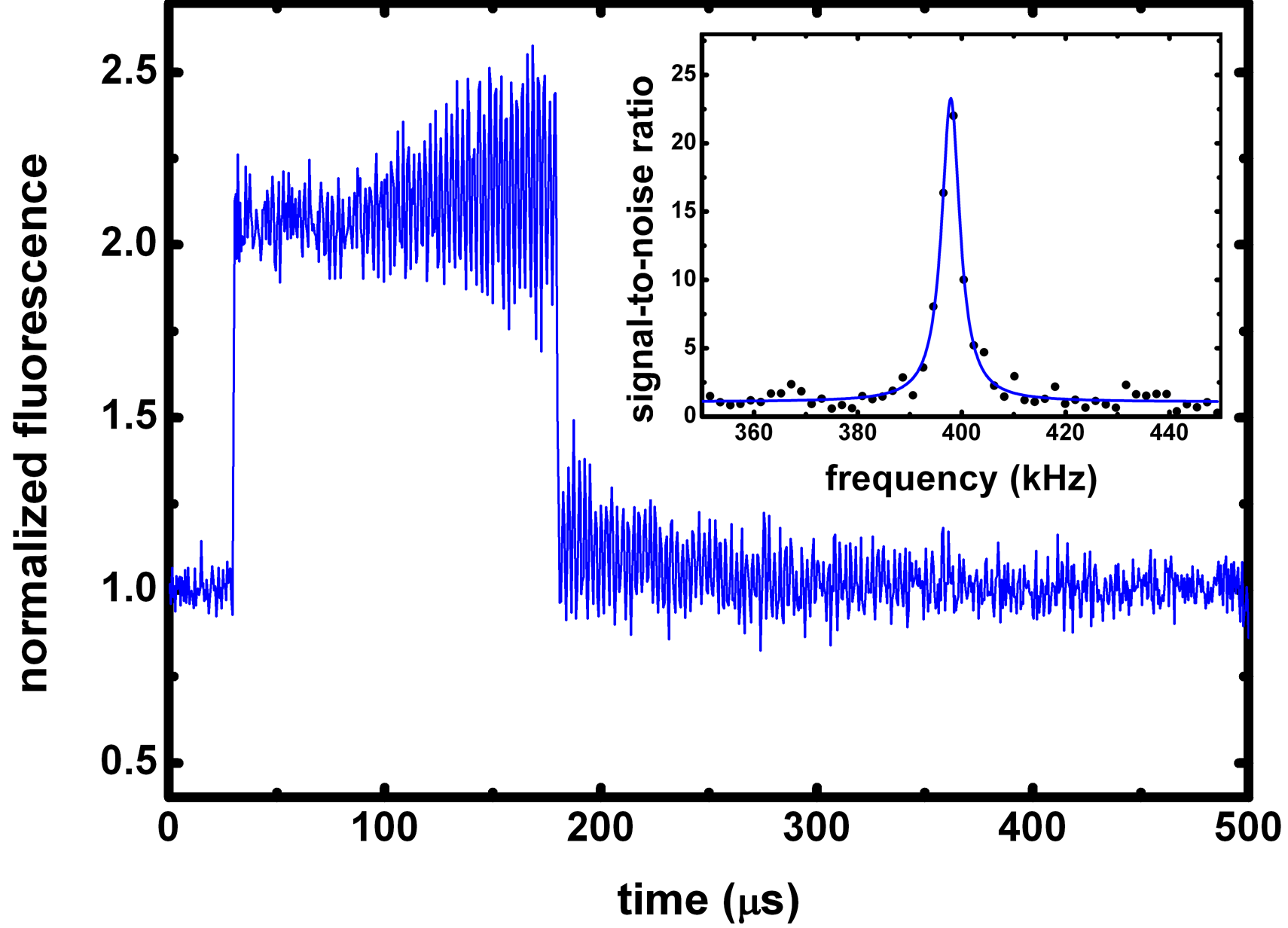}}
\caption{Fluorescence as a function of time elapsed since the experimental trigger produced using the laser frequency switching technique. The inset shows the FFT of the fluorescence profile with high contrast resonance peak at the COM-mode frequency. A Lorentzian function has been fit to the data.}
\label{fig:exp_detuning_shift_kick}
\end{center}
\end{figure}
\begin{figure}[b]
\begin{center}
\resizebox{0.475\textwidth}{!}{\includegraphics{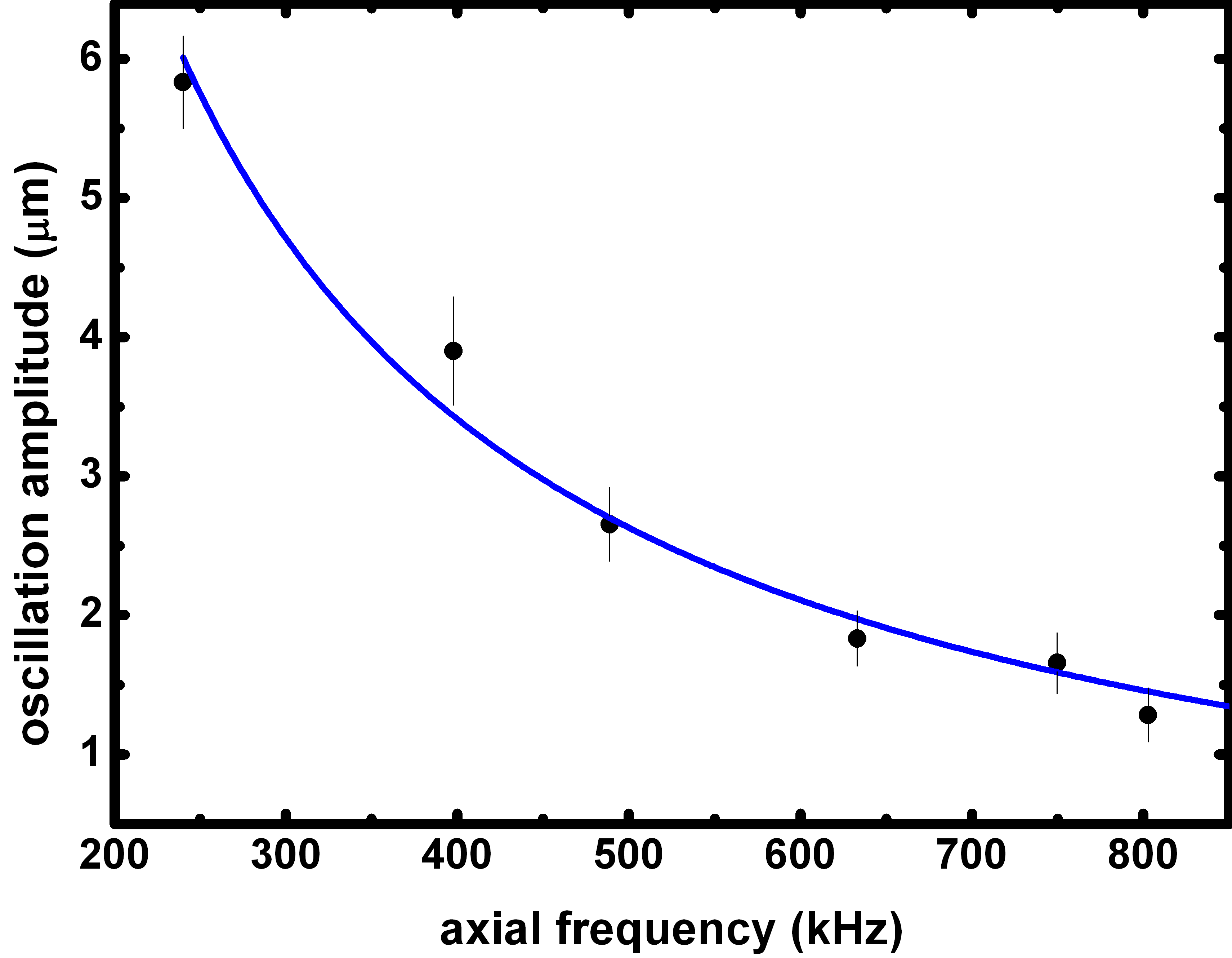}}
\caption{The ion's oscillation amplitude as a function of the axial confinement. For each measurement the laser detuning is +10~MHz during amplification and -40~MHz during re-cooling. The laser intensity is $I_{sat}$ and the amplification time is 100~$\mu$s. The x-error bars are smaller than the data points. The expected reciprocal function has been fit to the data.}
\label{fig:exp_axial_range}
\end{center}
\end{figure}
An optical kick can be imparted to the ion by switching between two laser detunings provided that there is a difference in the magnitude of the radiation pressure. When the laser is rapidly shifted from red to blue detuning, the ion's equilibrium position in the trapping potential is shifted and the resulting oscillatory motion is coherently amplified through interaction with the laser. The phase of the ion's motion at the switching time becomes negligible since the amplified oscillation is large compared to the oscillation induced by the light pressure shift. Thus, restrictions on the timing of the laser sequence are removed.

The excitation in this scheme can be modeled by a step function where the excitation amplitude is proportional to the inverse secular frequency. This technique is particularly suited for situations where the secular frequency is unknown or the frequency is expected to experience large changes during the course of an experiment. However, due to the relatively large difference between red and blue detunings necessary for supplying the optical excitation, the overall fluorescence signal level may be lower than for the laser intensity switching procedure described in section \ref{sec:exp_procedure}.

Figure \ref{fig:exp_detuning_shift_kick} shows a fluorescence profile for a laser sequence where the optical excitation is delivered through a shift in the laser detuning. The laser is initially red detuned for 30~$\mu$s to cool the ion's motion. The laser frequency is then suddenly shifted to small blue detuning ($\delta\!\! =\!\! +\,10$~MHz) with respect to the red detuning ($\delta\!\! =\!\! -\,40$~MHz) and the ion's oscillatory motion is amplified for 150~$\mu$s. The difference in radiation pressure between the red and blue detunings determines the magnitude of the excitation $\Delta F$ (see equation \ref{equ:F_mod}). Following amplification, the laser is once again red detuned leading to damping of the ion's motion and restoring its temperature to the equilibrium value. This sequence is repeated at a rate of 1~kHz. The inset shows the ion's motional spectrum obtained by taking the FFT of the fluorescence profile. The motional spectrum contains a high contrast peak centered at the COM-mode frequency. 

We have measured the ion's oscillation amplitude for the entire range of available axial confinements while keeping the frequency switching parameters constant. The result is shown in figure \ref{fig:exp_axial_range} and the expected reciprocal function has been fit to the data with~good agreement.

%%%%%%%%%%%%%%%%%%%%%%%%%%%%%%%%%%%%%%%%%%%%%%%%%%%%%%%%%%%%%%%%%%%%%%%%%%%%%%%%%%%%%%%%%%%%%%%%%%%%%%%%%%%%%%%%%%%%%%%%%%%%%%%%%%%%%%%%%%
%%CONCLUSION%%
%%%%%%%%%%%%%%%%
\section{\label{sec:conclusion}Conclusion}
We have demonstrated and characterized a novel technique for the optical broadband excitation of the secular motion of trapped ions. Employing a blue detuned laser, small light pressure forces are coherently amplified until the ion achieves the desired amplitude of oscillatory motion. We have used a molecular dynamics simulation to model the system and characterize its parameters. We have determined the measurement induced temperature increase of the ion as a function of the laser amplification and demonstrated that it can be kept below 10~mK making the scheme well-suited for the interrogation of large 3D ion crystals.

This technique is a valuable tool for experimental applications where an electronic excitation of the secular motion is undesirable or inconvenient. High precision frequency measurements can be achieved within interrogation times on the order of seconds where the only requirement is laser line-of-sight. In addition, since the excitation has a reciprocal dependence on the secular frequency, the ion's COM-mode motion can be excited and amplified over a broad range of confinement potentials. 

In conclusion, we have investigated and characterized a novel optical broadband excitation technique which allows for the precisely controlled excitation and amplification of the ion's secular motion with a single interaction laser.

%%%%%%%%%%%%%%%%%%%%%%%%%%%%%%%%%%%%%%%%%%%%%%%%%%%%%%%%%%%%%%%%%%%%%%%%%%%%%%%%%%%%%%%%%%%%%%%%%%%%%%%%%%%%%%%%%%%%%%%%%%%%%%%%%%%%%%%%%%
%%ACKNOWLEDGMENTS%%
%%%%%%%%%%%%%%%%
\begin{acknowledgments}
This work was supported by the Engineering and Physical Sciences Research Council (EPSRC) of the UK.
\end{acknowledgments}

\end{document}